\documentstyle[preprint,aps]{revtex}

\def\singlespace {\smallskipamount=3.75pt plus1pt minus1pt
                  \medskipamount=7.5pt plus2pt minus2pt
                  \bigskipamount=15pt plus4pt minus4pt
                  \normalbaselineskip=12pt plus0pt minus0pt
                  \normallineskip=1pt
                  \normallineskiplimit=0pt
                  \jot=3.75pt
                  {\def\smallskip {\vskip\smallskipamount}}
                  {\def\medskip   {\vskip\medskipamount}}
                  {\def\bigskip   {\vskip\bigskipamount}}
                  {\setbox\strutbox=\hbox{\vrule
                    height10.5pt depth4.5pt width 0pt}}
                  \parskip 7.5pt
                  \normalbaselines}
\def\middlespace {\smallskipamount=5.625pt plus1.5pt minus1.5pt
                  \medskipamount=11.25pt plus3pt minus3pt
                  \bigskipamount=22.5pt plus6pt minus6pt
                  \normalbaselineskip=22.5pt plus0pt minus0pt
                  \normallineskip=1pt
                  \normallineskiplimit=0pt
                  \jot=5.625pt
                  {\def\smallskip {\vskip\smallskipamount}}
                  {\def\medskip   {\vskip\medskipamount}}
                  {\def\bigskip   {\vskip\bigskipamount}}
                  {\setbox\strutbox=\hbox{\vrule
                    height15.75pt depth6.75pt width 0pt}}
                  \parskip 11.25pt
                  \normalbaselines}
\def\doublespace {\smallskipamount=7.5pt plus2pt minus2pt
                  \medskipamount=15pt plus4pt minus4pt
                  \bigskipamount=30pt plus8pt minus8pt
                  \normalbaselineskip=30pt plus0pt minus0pt
                  \normallineskip=2pt
                  \normallineskiplimit=0pt
                  \jot=7.5pt
                  {\def\smallskip {\vskip\smallskipamount}}
                  {\def\medskip   {\vskip\medskipamount}}
                  {\def\bigskip   {\vskip\bigskipamount}}
                  {\setbox\strutbox=\hbox{\vrule
                    height21.0pt depth9.0pt width 0pt}}
                  \parskip 15.0pt
                  \normalbaselines}
\def\al{\alpha}

\def\th{\theta}

\def\si{\sigma}

\def\ph{\phi}

\def\Ph{\Phi}

\def\cF{{\cal F}}
\def\cL{{\cal L}}

\def\frac#1#2{\textstyle{{{#1} \over {#2}}}}

\def\prt{\partial}

\def\half{{\textstyle{1\over 2}}}
\def\lsim{\mathrel{\rlap{\lower4pt\hbox{\hskip1pt$\sim$}}
    \raise1pt\hbox{$<$}}}
\def\gsim{\mathrel{\rlap{\lower4pt\hbox{\hskip1pt$\sim$}}
    \raise1pt\hbox{$>$}}}

\newcommand{\beq}{\begin{equation}}
\newcommand{\eeq}{\end{equation}}
\newcommand{\bea}{\begin{eqnarray}}
\newcommand{\eea}{\end{eqnarray}}
\newcommand{\rf}[1]{(\ref{#1})}

\tightenlines
\begin{document}
\preprint{
\hfill$\vcenter{\hbox{\bf IUHET-462} \hbox{December
             2003}}$  }

\title{\vspace*{.75in}
Superfield Realizations of Lorentz Violation\footnote{Presented at 
the Third International Symposium on Quantum Theory and Symmetries, 
10-14 September 2003, Cincinnati, OH}}

\author{M. S. Berger
\footnote{Electronic address:
berger@indiana.edu}}

\address{
Physics Department, Indiana University, Bloomington, IN 47405, USA}

\maketitle

\thispagestyle{empty}

\begin{abstract}
Lorentz-violating extensions of the Wess-Zumino model have been formulated in
superspace. The models respect a supersymmetry algebra and can be 
understood as arising from suitably modified superspace transformations.
\end{abstract}

\newpage

\section{Introduction}

Spacetime symmetries have always played a central role in
particle physics. Some of these symmetries
like the Poincar\'e symmetry are taken to be exact, while other symmetries
like supersymmetry are assumed to be broken. From the perspective of 
experimental physics it appears that supersymmetry is a badly broken symmetry, 
so much so that there has to date been no direct evidence for it. However
viewed from the Planck scale, electroweak-scale 
supersymmetry is almost exact. Indeed much theoretical research has been 
devoted to accounting for the ratio of the supersymmetry breaking scale to the 
Planck scale. From this point of view, it seems that if one accepts the 
possibility that the a spacetime symmetry such as supersymmetry is broken 
in a very small manner, that one should entertain the possibility that 
the remaining spacetime symmetries are also broken to a very small extent 
even though there is no experimental evidence for it.

As the possibility that Lorentz and CPT violation might occur in fundamental
theories has become more apparent, there has been interest 
in finding supersymmetric theories 
with Lorentz 
violation\cite{Berger:2001rm,Berger:2003ay,Belich:2003fa}.
One approach to incorporate Lorentz and CPT violation into a 
Lorentz-symmetric Lagrangian by adding explicit breaking 
terms\cite{Colladay:1996iz,Colladay:1998fq}.
The resulting field theories should be regarded as effective theories
arising from the more fundamental theory. 
Problems with microcausality are addressed in the underlying fundamental 
theory at the energy scales at which the effective theory breaks
down\cite{Kostelecky:2000mm}. The experimental implications of Lorentz and 
CPT violation parameterized in this manner have been explored extensively in 
recent years\cite{cpt01}.

The first supersymmetric model with Lorentz and CPT violation involved 
extending\cite{Berger:2001rm} the Wess-Zumino model\cite{Wess:tw}.
Two extensions were found, and these two models
admit a superspace formulation\cite{Berger:2003ay}.

\section{The Wess-Zumino Lagrangian}

A useful tool for developing supersymmetric field theories is to represent 
a supermultiplet of component fields as a superfield defined over a superspace
of coordinates
\bea
&&z^M=(x^\mu,\th ^\al,\bar\th _{\dot\al})
\eea
The four anticommuting coordinates $\th ^\al$ and $\bar\th _{\dot\al}$ 
form two-component Weyl spinors.
A superfield $\Ph(x,\th,\bar\th)$ is then
a function
of the commuting spacetime coordinates $x^\mu$ and of four anticommuting 
coordinates $\th ^\al$ and $\bar\th _{\dot\al}$ which form two-component Weyl 
spinors.
A chiral superfield is a function of 
$y^\mu =x^\mu+i\th \si ^\mu \bar \th$ and
$\th$, i.e.
\bea 
\Ph(x,\th,\bar\th)&=&\ph(y)+\sqrt{2}\th \psi(y)+(\th \th)\cF(y)\;, \nonumber \\
&=&\ph(x)+i\th \si ^\mu \bar\th\prt _\mu \ph(x)
-{1\over 4}(\th \th)(\bar\th \bar\th)\Box \ph(x)\nonumber \\
&&+\sqrt{2}\th \psi(x)+i\sqrt{2}\th \si ^\mu \bar\th \th\prt _\mu \psi(x)
+(\th \th)\cF(x)\;,
\eea
where one can define the usual real components of the complex scalar components
as
\bea
&&\ph = \frac 1 {\sqrt 2} (A + i B),
\quad
\cF = \frac 1 {\sqrt 2} (F - i G).
\eea
The conjugate superfield is 
\bea 
\Ph^*(x,\th,\bar\th)&=&\ph^*(z)+\sqrt{2}\bar\th \bar\psi(z)
+(\bar\th \bar\th)\cF^*(z)
\;, 
\eea
where $z^\mu=y^{\mu *}=x^\mu -i\th \si ^\mu \bar \th$.
The Wess-Zumino Lagrangian can now be derived from the superspace 
integral\cite{Salam:1974yz}
\bea
\int d^4\th \Ph^*\Ph + \int d^2\th \left [ 
{1\over 2}m\Ph^2 +{1\over 3}g\Ph^3 
+h.c.\right ]\;.
\label{superspace}
\eea
The superspace integral over $\int d^4\th$ projects out the 
$(\th \th)(\bar \th \bar\th)$ component of the $\Ph^* \Ph$ 
superfield while the $\int d^2\th$ projects out the 
$\th \th$ component of the superpotential.
The result 
\bea
\cL_{WZ}&=&\prt _\mu \phi^* \prt ^\mu \ph 
+{i\over 2}[(\prt _\mu \psi) \si ^\mu 
\bar\psi+(\prt _\mu\bar\psi)\bar\si ^\mu \psi]
+\cF^*\cF \nonumber \\
&&+m\left [\ph \cF + \ph ^*\cF^* -\half\psi\psi -\half\bar\psi\bar\psi\right ] 
\nonumber \\
&&+g\left [\ph^2\cF+\ph^{*2}\cF^*-\ph (\psi\psi)-\ph^*(\bar\psi\bar\psi)
\right ]\;,
\label{wz}
\eea
is a Lagrangian
which transforms into itself plus a total derivative under a supersymmetric 
transformation.

One can formulate the Wess-Zumino model in terms of differential operators that
act on the superfields. Define
\bea
&&X\equiv (\th \sigma ^\mu \bar\th)\prt _\mu\;, 
\eea
so that
\bea
&&U_x \equiv e^{iX}=1+i(\th \sigma ^\mu \bar\th)\prt _\mu -{1\over 4}(\th \th)
(\bar\th \bar\th)\Box \;. 
\eea
The expansion terminates because of the anticommuting nature of $\th$ and 
$\bar\th$.
Since $X$ is a derivative operator, the action of $U_x$
on a superfield ${\mathcal S}$ is a coordinate shift in which the 
spacetime coordinate $x^\mu$ is shifted to $y^\mu$,
\bea
&&U_x{\mathcal S}(x,\th,\bar\th)={\mathcal S}(y,\th,\bar\th)\;.
\eea
The chiral superfield $\Ph(x,\th,\bar\th)$ is a function of $y^\mu$ and 
$\th$ only, so it must then be of the form $\Ph(x,\th,\bar\th)=U_x\Psi(x,\th)$
for some function $\Psi$. The kinetic terms of the 
Wess-Zumino model can be expressed as 
\bea
&&\int d^4\th \left [U_x^*\Psi(x,\bar\th)^*\right ]\left [U_x\Psi(x,\th)
\right ]
=\int d^4\th \Ph^*(z,\bar\th)\Ph(y,\th)\;.
\eea

\section{Lorentz Violation}

It has been shown on quite general grounds
that CPT violation implies 
Lorentz violation\cite{Greenberg:2002uu,Greenberg:2003ks}, 
but it is possible to 
have Lorentz violation without CPT violation..
Of the two extensions to the Wess-Zumino model, 
the first does not contain CPT violation and it
is easily understood as arising from the 
following substitution
$\prt _\mu \to \prt _\mu +k_{\mu\nu}\prt ^\nu$, where $k_{\mu\nu}$ is a real, 
symmetric, traceless, and dimensionless coefficient 
responsible for the Lorentz violation.
so that
\bea
\cL_{\rm Lorentz}&=&\cL_{WZ}(\prt _\mu \to \prt _\mu +k_{\mu\nu}\prt ^\nu)\;.
\label{Lorentz}
\eea
The coefficient $k_{\mu\nu}$
transforms as a 2-tensor under observer Lorentz transformations
but as a scalar under particle Lorentz transformations\footnote{The terms 
containing $k_{\mu\nu}$ give rise to Lorentz violation,
the physics remains independent of the particular coordinate system that is 
used to describe it.}. Adding Lorentz violation in 
this fashion can be immediately extended to encompass supersymmetric gauge 
theories as well since the superfield formulations of those theories will
carry forward under the substitution 
$\prt _\mu \to \prt _\mu +k_{\mu\nu}\prt ^\nu$.

The second extension of the Wess-Zumino model violates CPT in addition to 
containing Lorentz violation and is given by the 
Lagrangian\cite{Berger:2001rm}
\bea
\cL_{\rm CPT}&=&\left [(\prt _\mu-ik_\mu) \phi^*\right ]
\left [ (\prt ^\mu +ik^\mu)\ph\right ]\nonumber \\ 
&&+{i\over 2}[((\prt _\mu +ik_\mu)\psi) \si ^\mu 
\bar\psi+((\prt _\mu-ik_\mu)\bar\psi)\bar\si ^\mu \psi]
+\cF^*\cF\;.
\label{CPT}
\eea
Here the Lorentz and CPT violation is controlled by $k_\mu$, which is a real 
coefficient of mass dimension one which transforms as a four-vector 
under observer 
Lorentz transformations but remains unaffected by particle Lorentz 
transformations\cite{Colladay:1996iz,Colladay:1998fq}. Unlike the coefficient
$k_{\mu \nu}$, the quantity $k_\mu$ has an odd number of four-indices so it
violates CPT. 
The Lagrangian for the model with the CPT-violating coefficient $k_\mu$ 
can be obtained from the kinetic part of the Wess-Zumino Lagrangian in 
Eqn.~\rf{wz} with the appropriate substitutions 
$\prt _\mu \to \prt _\mu \pm ik_\mu$.

The two Lorentz-violating models can be described in the superspace 
formalism in a way that parallels that of the ordinary Wess-Zumino model.
Define superfields
\bea 
\Ph_y(x,\th,\bar\th)
&=&\Ph(x,\th,\bar\th;\prt _\mu \to \prt _\mu +k_{\mu\nu}\prt ^\nu)
\nonumber \\
&=&\ph(x_+)+\sqrt{2}\th \psi(x_+)+(\th \th)F(x_+)\;, 
\eea
and 
\bea 
\Ph^*_y(x,\th,\bar\th)
&=&\Ph^*(x,\th,\bar\th;\prt _\mu \to \prt _\mu +k_{\mu\nu}\prt ^\nu)
\nonumber \\
&=&\ph^*(x_-)+\sqrt{2}\bar\th \bar\psi(x_-)+(\bar\th \bar\th)F^*(x_-)
\;, 
\eea
where 
\bea
&&x_\pm^\mu =x^\mu \pm i\th \si ^\mu \bar\th \pm ik^{\mu \nu}\th 
\si_\nu\bar\th\;.,
\eea
are shifted coordinates that take the place of $y^\mu$ and $z^\mu$.
Under a CPT-transformation the chiral superfield $\Ph_y$ and the 
antichiral superfield $\Ph^*_y$ transform into themselves just as the usual
superfields $\Ph$ and $\Ph^*$ do. The Lagrangian in Eqn.~\rf{Lorentz} can 
be obtained by the same superspace integral in Eqn.~\rf{superspace} with 
the superfields $\Ph_y$ and $\Ph^*_y$ substituted in the place of 
$\Ph$ and $\Ph^*$ (see Eqn.~\rf{superspace2} below).

The appropriate modified superfields for the model in Eqn.~\rf{CPT},
\bea 
\Ph_k(x,\th,\bar\th)
&=&\Ph(x,\th,\bar\th;\prt _\mu \to \prt _\mu +ik_\mu)\;,
\label{CPTsubs1}
\eea
and
\bea 
\Ph^*_k(x,\th,\bar\th)
&=&\Ph^*(x,\th,\bar\th;\prt _\mu \to \prt _\mu -ik_\mu)\;,
\label{CPTsubs2}
\eea
cannot be obtained from a coordinate shift, but can be obtained from a more
general superspace transformation.


\section{Superspace Transformations}

The Lorentz-violating extensions of the Wess-Zumino
model can be understood as transformations on the superfields.
Define
\bea
&&Y\equiv k_{\mu\nu}(\th \sigma ^\mu \bar\th)\prt ^\nu\;, \\
&&K\equiv k_\mu(\th \sigma ^\mu \bar\th)\;,
\eea
so that
\bea
&&U_y \equiv e^{iY}=1+ik_{\mu\nu}(\th \sigma ^\mu \bar\th)\prt^\nu
 -{1\over 4}k_{\mu \nu}k^{\mu \rho}(\th \th)
(\bar\th \bar\th)\prt ^\nu \prt _\rho\;, \\
&&T_k \equiv e^{-K}=1-k_\mu(\th \sigma ^\mu \bar\th)+{k^2\over 4}(\th \th)
(\bar\th \bar\th)\;.
\eea
Since $Y$, like $X$, is a derivative operator, the action of $U_y$
on a superfield ${\mathcal S}$
is a coordinate shift. The appearance of terms of 
order ${\mathcal O}(k^2)$ in both cases is easily understood in terms of 
these operators.
Furthermore we have $U_y^*=U_y^{-1}$ 
while $T_k^*=T_k$ and not its inverse.

The supersymmetric models with Lorentz-violating terms 
can be expressed in terms of new superfields,
\bea
\Ph_y(x,\th,\bar\th)&=&U_yU_x\Psi(x,\th)\;, \\
\Ph^*_y(x,\th,\bar\th)
&=&U_y^{-1}U_x^{-1}\Psi^*(x,\bar\th)\;.
\eea
Applying $U_y$ to the chiral and antichiral superfields merely effects the
substitution $\prt _\mu \to \prt _\mu +k_{\mu\nu}\prt ^\nu$. Since $U_y$ 
involves a derivative operator just as $U_x$, the derivation of the chiral 
superfield $\Ph_y$ is a function of the variables $x_+^\mu$
and $\theta$ analogous to how, in the usual case, $\Ph$ is a function 
of the variables $y^\mu$ and $\theta$. The 
Lagrangian is given by 
\bea
&&\int d^4\th \Ph_y^*\Ph_y+ \int d^2\th \left [ 
{1\over 2}m\Ph_y^2 +{1\over 3}g\Ph_y^3 
+h.c.\right ]\nonumber \\
&&=\int d^4\th \left [U_y^*\Ph^*\right ]
\left [U_y\Ph\right ]+ \int d^2\th \left [ 
{1\over 2}m\Ph^2 +{1\over 3}g\Ph^3 
+h.c.\right ]\;.
\label{superspace2}
\eea

For the CPT-violating model the superfields have the form
\bea
\Ph_k(x,\th,\bar\th)&=&T_kU_x\Psi(x,\th)\;, \\
\Ph^*_k(x,\th,\bar\th) 
&=&T_kU_x^{-1}\Psi^*(x,\bar\th)\;.
\eea
It is helpful to note that the transformation $U_x$ acts on $\Psi$ and 
its inverse $U_x^{-1}$ 
acts on $\Psi^*$, while the same transformation $T_k$ acts on both 
$\Psi$ and $\Psi^*$ (since $T_k^*=T_k$). A consequence of this fact is 
that the supersymmetry transformation will act differently on the components 
of the chiral superfield and its conjugate. 
Specifically the chiral superfield $\Ph_k$ is the
same as $\Ph$ with the substitution $\prt _\mu \to \prt _\mu +ik_\mu$ 
whereas the 
antichiral superfield $\Ph_k^*$ is the same as $\Ph^*$ with the 
substitution $\prt _\mu \to \prt _\mu -ik_\mu$, as in Eqns.~\rf{CPTsubs1}
and \rf{CPTsubs2}.

The CPT-violating model in Eqn.~\rf{CPT} 
can then be represented in the following way as a 
superspace integral:
\bea
&&\int d^4\th \Ph_k^*\Ph_k=\int d^4\th \Ph^* e^{-2K}\Ph
\label{proj}
\eea
Unlike the CPT-conserving model, the $(\th \th)(\bar\th \bar\th)$ component
of $\Ph^*\Ph$ no longer transforms into a total derivative. A specific  
combination of components of $\Ph^*\Ph$ does transform into a total 
derivative, and this combination is in fact the 
$(\th \th)(\bar\th \bar\th)$ component of $\Ph_k^*\Ph_k$.


\section{Conclusions}

The Wess-Zumino model can be described in 
terms of superspace transformations and the projections of
components of functions of the superfield.
It was shown that the two Lorentz-violating extensions of the Wess-Zumino
model can be understood 
in terms of analogous transformations on  
modified superfields and projections arising from superspace integrals.


\section*{Acknowledgments}

This work was supported in part by the U.S.
Department of Energy under Grant No. No.~DE-FG02-91ER40661.





\end{document}